\documentclass[epj]{webofc}
\usepackage[varg]{txfonts}   % Web of Conferences font
\usepackage{color}
\allowdisplaybreaks[1]
\wocname{EPJ Web of Conferences}
\woctitle{New Frontiers in Physics 2014}
\begin{document}
%%%%% PREPRINT NUMBERS %%%%%%
\begin{flushright}
%28/04/2005
KIAS-P14072\\
KOBE-TH-14-13\\
OU-HET-839%TU-793
%hep-th/0305xxx
\end{flushright}
\vspace{-1.5cm}
\selectlanguage{english}
\title{Realization of the Lepton Flavor Structure from Point Interactions}
%
% subtitle is optional

\author{Yukihiro FUJIMOTO\inst{1}\fnsep\thanks{\email{fujimoto@het.phys.sci.osaka-u.ac.jp}} \and
        Kenji NISHIWAKI\inst{2}\inst{3} \and
        Makoto SAKAMOTO\inst{4} \and
        Ryo TAKAHASHI\inst{5}
        % etc.
}

\institute{Department of Physics, Osaka University, Machikaneyama-Cho 1-1, Toyonaka 560-0043, Japan
\and
           {Regional Centre for Accelerator-based Particle Physics, Harish-Chandra Research Institute,
Allahabad 211 019, India}
\and
	{School of Physics, Korea Institute for Advanced Study, 85 Hoegiro, Dongdaemun-gu, Seoul, 130-722 Republic of Korea}
\and
          Department of Physics, Kobe University,
1-1 Rokkodai, Nada, Kobe 657-8501, Japan
\and
	Graduate School of Science and Engineering, Shimane University,
1060 Nishikawatsu, Matsue, Shimane 690-8504, Japan
          }

\abstract{%
  We investigate {a} 5d gauge theory on $S^1$ with point interactions. The point interactions describe extra boundary conditions and provide three generations, the charged lepton mass hierarchy, the lepton flavor mixing and tiny degenerated neutrino masses after choosing suitable boundary conditions and parameters. The existence of the restriction {in} the flavor mixing, which appears from the configuration of the extra dimension, is one of the {features} of this model. Tiny Yukawa couplings for the neutrinos also appears without the see-saw mechanism nor symmetries in our model. {The magnitude of CP violation in the leptons} can be a prediction and is consistent with the current experimental data.
}
\maketitle
\section{Introduction}
\label{intro}
The Standard Model (SM) has {done} well until today. However, it does not mean that there is no problem. {Even though} a SM-like Higgs was found, the problems {relating} to the Yukawa couplings are left. The generation problem is one of them. Both in the quark sector and the lepton sector, we have the structure of the three generations. Every generation {has exactly} the same quantum numbers except for their masses. There is only {a} phenomenological reason to introduce the three generations, which is for the CP phase~\cite{Kobayashi:1973fv}, but there is no theoretical {reason} to them. The mass hierarchy is another one. There exists a large mass hierarchy for the quarks and the charged leptons. In the SM, the Yukawa couplings are just parameters and there is no reason for producing a large mass hierarchy. {{Tininess in the neutrino masses }is also another problem. From experimental data, it had been 
found that {the differences in the squared neutrino masses} are of order ${\cal O}(10^{-4}-10^{-3})$\ {eV}. The origin of the tiny neutrino masses is still unknown.} The flavor mixing structure of the quark sector and the lepton sector, which {is} controlled by the Cabbibo-Kobayashi-Maskawa (CKM) matrix and {the} Pontecorvo-Maki-Nakagawa-Sakata (PMNS) matrix in the SM, is also a mystery. {Due to} the improvement of {experiments}, every mixing {angle is} measured with the CP phase {in} the quark sector, however, it is still unknown what kind of dynamics controls {these} values. Not only the above problems but also many other problems {remain} in the SM. 

A way to solve the above problems {in} the SM is to use extra dimensions. Especially, 5d {gauge} theories with point interactions is an attractive one for solving the generation problem and the mass hierarchy problem {in} the fermions with producing the realistic flavor mixing angle from the extra dimension~\cite{Fujimoto:2012wv,Fujimoto:2014fka}. {In the paper~\cite{Fujimoto:2012wv}, the generation problem and the mass hierarchy problem in the quark sector are discussed and are solved with the realistic flavor mixing and the CP phase. In the research~\cite{Fujimoto:2014fka}, it turns out that we can apply the same method to the lepton sector for solving the generation problem and the mass hierarchy problem in the charged leptons with producing tiny neutrino masses and the realistic flavor mixing.} 
The point interaction is a generalized zero-width interaction, an example of which is a delta-function potential in 1d quantum mechanics (QM)~\cite{Reed:1975uy,Seva1986,Cheon:2000tq}, and is a key ingredient to produce the degenerated zero-mode functions, which is nothing but generations. By using the point interactions, we can put extra {boundary conditions (BC's)} for the fermions at that points as in the case of the delta-function potential in 1d QM. Because of this (extra) BC's, we can produce triply-degenerated zero-mode functions, which is nothing but three generations, without breaking the 5d gauge invariance. Since each triply-degenerated mode function lives in {a} different part of the extra dimension with {localization toward} the (extra) boundary points, we can produce a large hierarchy for the masses of the fermions through the overlap integrals if we introduce a 5d gauge-singlet scalar, which possesses {an} extra-dimensional coordinate-dependent vacuum expectation value (VEV) due to the {BC's}. The tiny degenerated neutrino masses are also realized in this model. The magnitude of the localization of the zero-mode functions is controlled by the bulk masses of the 5d fermions. To realize the tiny neutrino masses, we {should} produce the {extremely} tight localization with {setting} the values of the bulk masses {large}. Because of this tight localization, the effect of the gauge-single scalar VEV becomes {small,} so that a hierarchical mass structure does not appear {in} the neutrinos. {A characteristic} feature of this model is that the flavor mixing is restricted by the geometry of the extra dimension. Because of {this fact}, we cannot fill up all the elements of the mass matrix but only 6 we can. Thus it is non-trivial whether we can reproduce {experimental} values {in} the model {though} we have more parameters than the Standard Model. Furthermore, {the magnitude of CP violation in the lepton sector} can be a prediction in this model because we have only one source for the CP phase, which is nothing but a Higgs VEV~\cite{Fujimoto:2013ki}. {In our model, the Higgs VEV has a extra-dimensional coordinate-dependent phase with containing a twist parameter, which appears to the BC's for the Higgs. This extra-dimensional coordinate-dependent phase is the only source of the CP phase. When we treat the quark sector and the lepton sector at the same time, the twist parameter is used to fix the magnitude of CP violation in the quark sector so that the magnitude of CP violation in the lepton sector is automatically {determined~\cite{Fujimoto:2014fka}.}
}%{In our model,  } The VEV of the Higgs should be shared with both the quark sector and the lepton sector so that after we fixed the CP phase of the quark sector, there is no extra degree of freedom for the CP phase of the lepton sector~\cite{Fujimoto:2014fka}.

%As is shown in the next section, we can drive boundary conditions (BC's) for the fields from the point interaction method, which is developed in the researches of the point interaction in 1d QM. The BC's are important for the low energy effective theory because the zero modes are really sensitive to their BC's. Amazingly, by using 5d gauge theories with point interactions, we can solve the above problems of the SM as shown in the latter half.

\section{Lepton flavor structure from point interactions}
\label{LFSfromPI}

\subsection{A model}
\label{model}
\begin{figure}[h]
	\centering
	\sidecaption
	\includegraphics[width=9cm,clip]{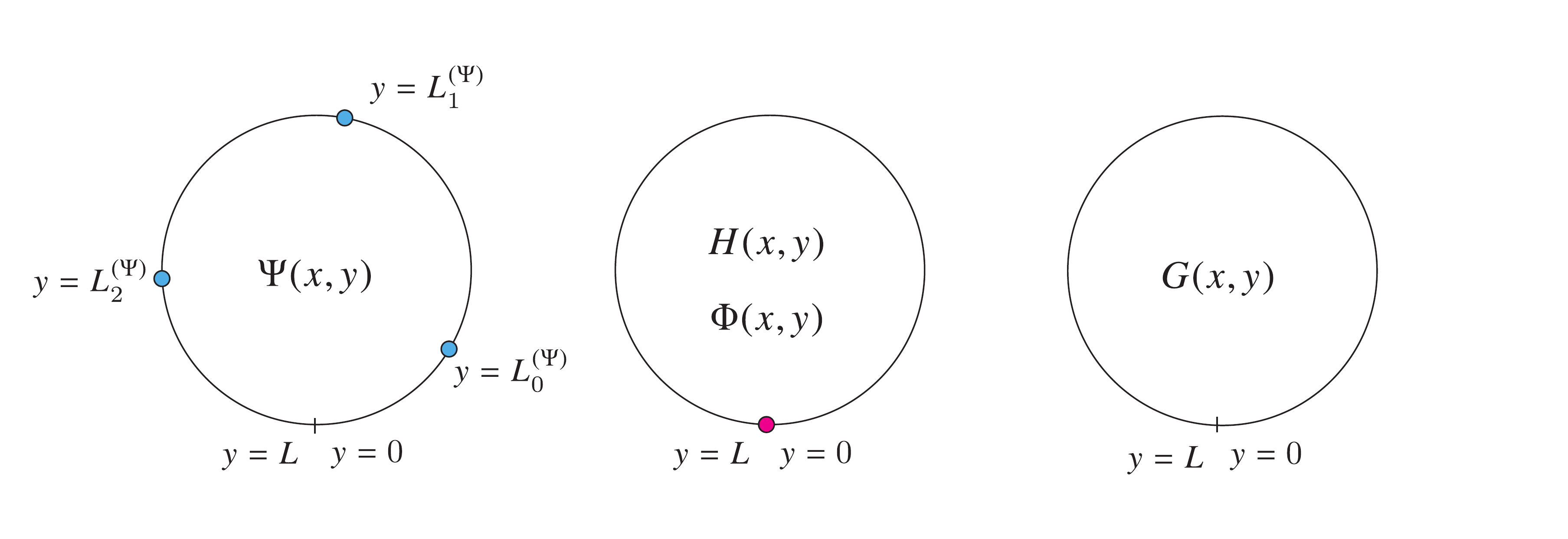}
	\caption{A schematic figure of the configuration of the point interactions. The fermions feel three point interactions, the gauge singlet scalar and the Higgs doublet feel one, the gauge fields do not feel the point interactions in the model.}
	\label{figure-configuration}       % Give a unique label
	\end{figure}

Our model consists of the following action.
	%%%%%%%%%%%%%%%%%%%%%%%%%%%%%%%%%%%%%%%%
	\begin{align}
	&S=S_{{\rm lepton}}+S_{{\rm Higgs}}+S_{{\rm singlet}}+S_{{\rm Yukawa}}^{({\rm lepton})},\label{action}\\[4pt]
		%%%
	&S_{{\rm lepton}}=\int d^4 x \int^{L}_{0}dy \left[\overline{L}(x,y)\Bigl(i\Gamma^{N}D_{N}^{(L)}+M_{L}\Bigr)L(x,y)\right.\nonumber\\
	&\left.\hspace{3em}+\overline{{\cal N}}(x,y)\Bigl( i\Gamma^{N} {\partial_{N}}+M_{{\cal N}}\Bigr) {\cal N}(x,y) +\overline{E}(x,y)\Bigl( i\Gamma^{N}D_{N}^{(E)}+M_{ E}\Bigr) E(x,y)\right],\label{Slepton}\\
		%%%
	&S_{{\rm Higgs}}=\int d^4 x \int^{L}_{0}dy \left[H^{\dagger}(x,y)\Bigl(D^{N}D_{N}+M^2\Bigr)H(x,y)-\frac{\lambda}{2}\Bigl(H^{\dagger}(x,y)H(x,y)\Bigr)^2\right],\\[4pt]
		%%%
	&S_{{\rm singlet}}=\int d^4 x \int^{L}_{0}dy \left[\Phi^{\dagger}(x,y)\Bigl(\partial^{N}\partial_{N}-M_{\Phi}^2\Bigr)\Phi(x,y)-\frac{\lambda_{\Phi}}{2}\Bigl(\Phi^{\dagger}(x,y)\Phi(x,y)\Bigr)^2\right],\label{5daction_singlet} \\
		%%%
	&S_{{\rm Yukawa}}^{({\rm lepton})}=\int d^4 x\int^{L}_{0}dy \left[ \Phi\Bigl(-{\cal Y}^{({\cal N})}\overline{ L}(i\sigma_{2}H^{\ast}){\cal N}-{\cal Y}^{(E)}\overline{L}HE\Bigr)+({\rm h.c.})\right]{,}\label{leptonYukawa}
	\end{align}
	%%%%%%%%%%%%%%%%%%%%%%%%%%%%%%%%%%%%%%%%
where we denote $L(x,y)$ as an $SU(2)_{W}$ doublet lepton, ${\cal N}(x,y)$ as an $SU(2)_{W}$ singlet neutrino, $E(x,y)$ as an $SU(2)_{W}$ singlet charged lepton,  $H(x,y)$ as the Higgs doublet and $\Phi(x,y)$ as {a} gauge singlet scalar field, respectively. 
$M_{\Psi}\ (\Psi=L, {\cal N}, E)$ {in Eq.~(\ref{Slepton})} denotes {the} bulk mass for the {fermions}, which is important to control the mass hierarchy in our {model. We} put the {signs} as
	%%%%%%%%%%%%%%%%%%%%%%%%%%%%%%%%%%%%%%%%
	\begin{align}
	&M_{L}<0,\label{signbulkmassL}\\
	&M_{{\cal N}}>0,\label{signbulkmassN}\\ 
	&M_{E}<0,\label{signbulkmassE}
	\end{align}
	%%%%%%%%%%%%%%%%%%%%%%%%%%%%%%%%%%%%%%%%
for {our purpose. 
Here, we omit the action for the {$SU(3)_C \times SU(2)_W \times U(1)_Y$} gauge fields for the simplicity.

Note that the discrete symmetry, $H\rightarrow -H$, $\Phi\rightarrow -\Phi$, is introduced to forbid the terms $\bar{L}(i\sigma_{2}H^{\ast}){\cal N}$, $\bar{L}HE$, $\Phi\bar{L}L$, $\Phi\bar{{\cal N}}{\cal N}$, $\Phi\bar{E}E$. We also simply ignore the term $(H^{\dagger}H)(\Phi^{\dagger}\Phi)$ in this model.

In the case of {an} extra dimension scenario, not only the action but also {the BC's are} important. In our model, every fermion field ($L,\, {\cal N},\, E$) feels three point interactions at the positions $y=L_{i}^{(\Psi)}$ ($\Psi=L, {\cal N}, E; \ i=0,1,2,3$), respectively. On the other hand, the Higgs {doublet $H$} and the gauge singlet scalar {$\Phi$} feel one point interaction at {$y=0$,} and gauge fields do not feel any point interaction, where the usual periodic {BC} is chosen.}
The situation is depicted in {Fig.}~\ref{figure-configuration}.

	%%%%%%%%%%%%%		figure 	%%%%%%%%%%%%%%%%
	%%%%%%%%%%%%%%%%%%%%%%%%%%%%%%%%%%%%%%%%
Because of the point interactions, every field feels {BC's} at {its own} positions. In our model, we adopt the following BC's for the purpose.
	%%%%%%%%%%%%%%%%%%%%%%%%%%%%%%%%%%%%%%%%
	\begin{align}
	L_{R}&=0\hspace{3em}{\rm at}\ \ \  y=L_{0}^{(L)},L_{1}^{(L)},L_{2}^{(L)},L_{3}^{(L)}\ , \label{LBC}\\
	{\cal N}_{L}&=0\hspace{3em}{\rm at}\ \ \ y=L_{0}^{({\cal N})},L_{1}^{({\cal N})},L_{2}^{({\cal N})},L_{3}^{({\cal N})}\ ,\label{NBC}\\
	E_{L}&=0\hspace{3em}{\rm at}\ \ \ y=L_{0}^{(E)},L_{1}^{(E)},L_{2}^{(E)},L_{3}^{(E)}\ .\label{EBC}
	\end{align}
	%%%%%%%%%%%%%%%%%%%%%%%%%%%%%%%%%%%%%%%
	%%%%%%%%%%%%%%%%%%%%%%%%%%%%%%%%%%
	\begin{align}
	\left\{
	\begin{array}{l}
	\Phi(x,0)+L_{+}\partial_{y}\Phi(x,0)=0,\\
	\Phi(x,L)-L_{-}\partial_{y}\Phi(x,L)=0,
	\end{array}
	\right.
	\hspace{3em}(-\infty\leq L_{\pm}\leq +\infty),\label{PhiBC}
	\end{align}
	%%%%%%%%%%%%%%%%%%%%%%%%%%%%%%%%%%%%%%%%
	%%%%%%%%%%%%%%%%%%%%%%%%%%%%%%%%%%%%%%%%
	\begin{align}
	\left\{
	\begin{array}{l}
	H(x,L)=e^{i\theta}H(x,0){,}\\
	\partial_{y}H(x,L)=e^{i\theta}\partial_{y}H(x,0),
	\end{array}
	\right.
	\hspace{3em}(-\pi<\theta\leq\pi),\label{HBC}
	\end{align}
	%%%%%%%%%%%%%%%%%%%%%%%%%%%%%%%%%%%%%%%%
with the identification of $S^1$,
	%%%%%%%%%%%%%%%%%%%%%%%%%%%%%%%%%%%%%%%%
	\begin{align}
	L &\sim 0,\\
	L_{3}^{({\cal N})} &\sim L_{0}^{({\cal N})},\\
	L_{3}^{(L)} &\sim L_{0}^{(L)},\\
	L_{3}^{(E)} &\sim L_{0}^{(E)},
	\end{align}
	%%%%%%%%%%%%%%%%%%%%%%%%%%%%%%%%%%%%%%%
where the indices $R$ and $L$ of the fermions denote the 4d chirality {defined} as $\Psi_{R}\equiv \left(\frac{1+\gamma_{5}}{2}\right)\Psi,\ \Psi_{L}\equiv \left(\frac{1-\gamma_{5}}{2}\right)\Psi$ and $L_{\pm}$ {are parameters which describe }the Robin BC.
{A suitable} choice of the parameters $L_{\pm}$ and other parameters in Eq.~(\ref{5daction_singlet}) makes the form of the VEV of the gauge-singlet {hierarchical} along the {$y$-direction}~\cite{Fujimoto:2011kf,Fujimoto:2012wv}.
Such a situation is preferable for generating the large hierarchy in the lepton mass matrices. $\theta$ is a phase parameter which specifies the twisted BC, whose complex degree of freedom is just the origin of CP violation in our model. 
We should emphasize that {all the BC's} are consistent with the 5d gauge invariance. {In particular,} the BC's~(\ref{LBC})--(\ref{EBC}) do not break the 5d gauge symmetry since the BC's for the fermion are given by the Dirichlet BC, which is manifestly invariant under the 5d gauge transformation.

\subsection{Generations}
\label{generations}

\begin{figure}[h]
	\centering
	\begin{minipage}{14cm}
	\includegraphics[width=0.99\columnwidth]{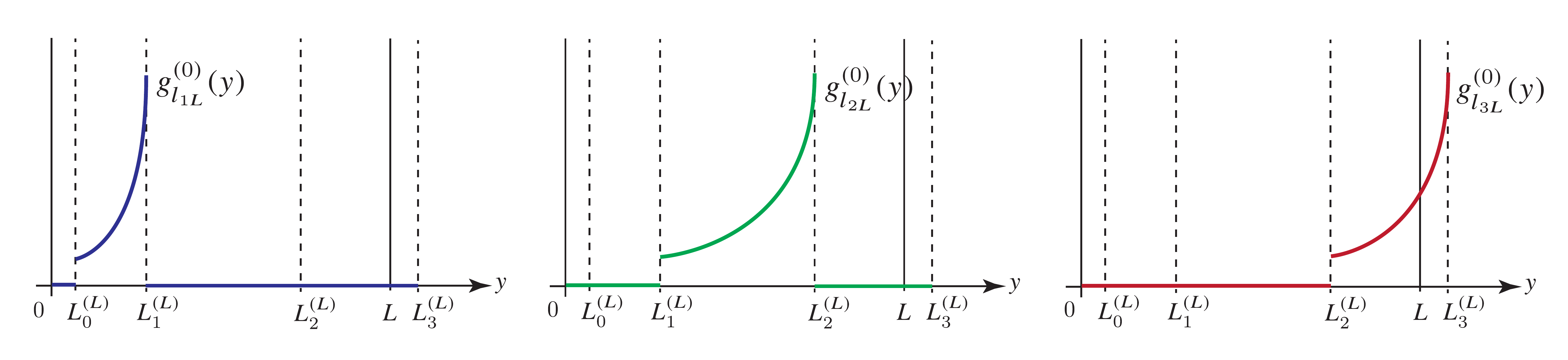}
	\par
		\vspace{0.0cm}
		{\footnotesize{(a) The profiles of triply-degenerated zero modes $g_{l_{iL}}^{(0)}$ with $M_{L}>0$. (The profiles of $f_{\nu_{iR}}^{(0)}$, $f_{e_{iR}}^{(0)}$ with $M_{\cal N}$, $M_{E}<0$.)}}
	\end{minipage}\\
	\begin{minipage}{14cm}
	\includegraphics[width=0.99\columnwidth]{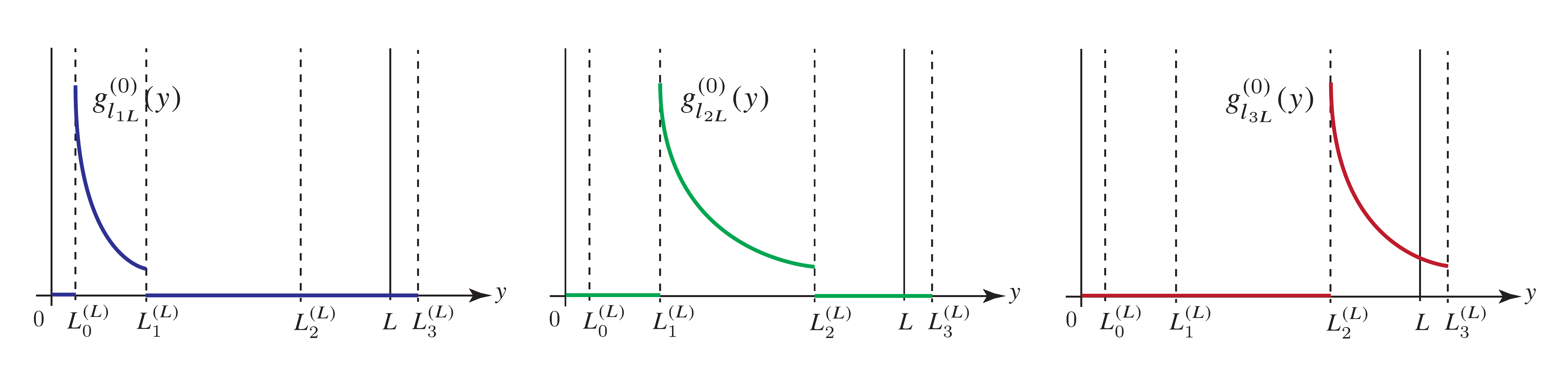}
	\par
		\vspace{0.0cm}
		{\footnotesize{(b) The profiles of triply-degenerated zero modes $g_{l_{iL}}^{(0)}$ with $M_{L}<0$. (The profiles of $f_{\nu_{iR}}^{(0)}$, $f_{e_{iR}}^{(0)}$ with $M_{\cal N}$, $M_{E}>0$.)}}
	\end{minipage}
	\caption{A schematic figure of the zero modes.}
	\label{figure-zeromodes}       % Give a unique label
	\end{figure}

Since we put the BC's~(\ref{LBC})--(\ref{EBC}) {on} the fermions with introducing the {point interactions}, three generations appears. The Kaluza-Klein (KK) expansion in this setup is done as
	%%%%%%%%%%%%%%%%%%%%%%%%%%%%%%%%%%%%%%%%
	\begin{align}
	\Psi(x,y)=\sum_{n}\left( \psi^{(n)}_{R}(x)f^{(n)}_{\psi_{R}}(y)+\psi^{(n)}_{L}(x)g_{\psi_{L}}^{(n)}\right), \label{KKexpansion}
	\end{align}
	%%%%%%%%%%%%%%%%%%%%%%%%%%%%%%%%%%%%%%%%
where $\{ f_{\psi_{R}}^{(n)}\}$ $\Bigl(\{ g_{\psi_{L}}^{(n)}\}\Bigr)$ is the eigenfunction of the hermitian operator $\mathcal{D}^{\dagger}\mathcal{D}$ $\left(\mathcal{D}\mathcal{D}^{\dagger}\right)$ and forms {a} complete set,
	%%%%%%%%%%%%%%%%%%%%%%%%%%%%%%%%%%%%%%%%
	\begin{align}
	\left\{
	\begin{array}{l}
	\mathcal{D}^{\dagger}\mathcal{D} f_{\psi_{R}}^{(n)}=m^{2}_{\psi^{(n)}}f_{\psi_{R}}^{(n)},\\[0.2cm]
	\mathcal{D}\mathcal{D}^{\dagger} g_{\psi_{L}}^{(n)}=m^{2}_{\psi^{(n)}}g_{\psi_{L}}^{(n)}{,}
	\end{array}
	\right. \hspace{3em}(\mathcal{D}\equiv \partial_{y}+M_{\Psi}, \ \mathcal{D}^{\dagger}\equiv -\partial_{y}+M_{\Psi}).
	\end{align}
	%%%%%%%%%%%%%%%%%%%%%%%%%%%%%%%%%%%%%%%%
{Zero mode solutions} of the above {should satisfy} the following equations, which {come} from quantum mechanical supersymmetry.
	%%%%%%%%%%%%%%%%%%%%%%%%%%%%%%%%%%%%%%%%
	\begin{align}
	\left\{
	\begin{array}{l}
	\mathcal{D} f_{\psi_{R}}^{(0)}=0,\\[0.2cm]
	\mathcal{D}^{\dagger} g_{\psi_{L}}^{(0)}=0.
	\end{array}
	\right.\label{zeromodeequations}
	\end{align}
	%%%%%%%%%%%%%%%%%%%%%%%%%%%%%%%%%%%%%%%%
Non-trivial solution of the above is 
	%%%%%%%%%%%%%%%%%%%%%%%%%%%%%%%%%%%%%%%%
	\begin{align}
	\left\{
	\begin{array}{l}
	f_{\psi_{R}}^{(0)}\propto e^{-M_{\Psi}y},\\[0.2cm]
	g_{\psi_{L}}^{(0)}\propto e^{+M_{\Psi}y}.
	\end{array}
	\right.
	\end{align}
	%%%%%%%%%%%%%%%%%%%%%%%%%%%%%%%%%%%%%%%%
Furthermore, we imposed the BC's in Eqs.~(\ref{LBC})-(\ref{EBC}) so that the three solutions appears to each segment. The situation is depicted in Fig \ref{figure-zeromodes}.

	%%%%%%%%%%%%%		figure 	%%%%%%%%%%%%%%%%
	%%%%%%%%%%%%%%%%%%%%%%%%%%%%%%%%%%%%%%

Thus, we {realize} three generations {in} chiral zero modes in our model.
	%%%%%%%%%%%%%%%%%%%%%%%%%%%%%%%%%%%%%%%%
	\begin{align}
	L(x,y)&=\sum_{i=1}^{3}l^{(0)}_{i L}(x)g_{l_{iL}}^{(0)}(y)+({\rm KK\ modes}),\\
	{\cal N}(x,y)&=\sum_{i=1}^{3}\nu^{(0)}_{i R}(x)f_{\nu_{iR}}^{(0)}(y)+({\rm KK\ modes}),\\
	E(x,y)&=\sum_{i=1}^{3}e^{(0)}_{i R}(x)f_{e_{iR}}^{(0)}(y)+({\rm KK\ modes}).
	\end{align}
	%%%%%%%%%%%%%%%%%%%%%%%%%%%%%%%%%%%%%%%%
Note that the mode functions are localized {toward} the boundary points because of the bulk mass $M_{\Psi}$ ($\Psi=L, {\cal N}, E$), the sign of whom determines the direction of {the} localization. We should note that the bulk mass controls all the related zero {modes,} so that we cannot change the form of the degenerated zero {modes} independently.

\subsection{Charged lepton mass hierarchy and tiny neutrino masses}
\label{masshierarchy}

Under the action~(\ref{action})--(\ref{leptonYukawa}) with the BC's~(\ref{LBC})--(\ref{HBC}), the lepton mass hierarchy and the tiny neutrino masses {appear}. {The corresponding mass terms are generated through the following part}.
	%%%%%%%%%%%%%%%%%%%%%%%%%%%%%%%%%%%%%%%%
	\begin{align}
	&S_{{\rm Yukawa}}^{({\rm lepton})}=\int d^4 x\int^{L}_{0}dy \left[ \Phi\Bigl(-{\cal Y}^{({\cal N})}\overline{ L}(i\sigma_{2}H^{\ast}){\cal N}-{\cal Y}^{(E)}\overline{L}HE\Bigr)+({\rm h.c.})\right]\nonumber\\
	&\quad {\supset -\int d^4 x \sum_{i,j=1}^{3}\left[ m_{ij}^{(\nu)} \overline{\nu_{iL}^{(0)}} \nu_{jR}^{(0)} + m_{ji}^{(\nu)}{}^{\ast} \overline{\nu_{jR}^{(0)}} \nu_{iL}^{(0)}\right]
- \int d^4 x \sum_{i,j=1}^{3}\left[ m_{ij}^{(e)} \overline{e_{iL}^{(0)}} e_{jR}^{(0)}
+ m_{ji}^{(e)}{}^{\ast} \overline{e_{jR}^{(0)}} e_{iL}^{(0)}\right]},
	\end{align}
	%%%%%%%%%%%%%%%%%%%%%%%%%%%%%%%%%%%%%%%%
where
	%%%%%%%%%%%%%%%%%%%%%%%%%%%%%%%%%%%%%%%%
	\begin{align}
	m_{ij}^{(\nu)}&={\cal Y}^{({\cal N})}\int^{L}_{0}dy \,\langle h(y)\rangle^{\ast} \,\langle \Phi(y) \rangle \,g_{l_{iL}}^{(0)}(y)\,f_{\nu_{jR}}^{(0)}(y),\label{nuoverlap}\\
	m_{ij}^{(e)}&={\cal Y}^{(E)}\int^{L}_{0}dy \,\langle h(y)\rangle \,\langle \Phi(y)\rangle \,g_{l_{iL}}^{(0)}(y)\,f_{e_{jR}}^{(0)}(y).\label{eoverlap}
	\end{align}
	%%%%%%%%%%%%%%%%%%%%%%%%%%%%%%%%%%%%%%%%
Under the Robin BC~(\ref{PhiBC}), it {is} known that the VEV of the gauge-singlet possesses the y-dependence and we can make it the exponential form by {choosing {suitable values} of $L_{\pm}$}, $M_{\Phi}$ and $\lambda_{\Phi}$~\cite{Fujimoto:2012wv,Fujimoto:2011kf}:
	%%%%%%%%%%%%%%%%%%%%%%%%%%%%%%%%%%%%%%%%
	\begin{align}
	\langle\Phi(y)\rangle=\phi(y)\sim e^{M_{\Phi}y}, \label{exponentialVEV}
	\end{align}
	%%%%%%%%%%%%%%%%%%%%%%%%%%%%%%%%%%%%%%%%
Obviously, the form of the VEV {in} Eq.~(\ref{exponentialVEV}) makes a big differences {in} the overlap integrals (\ref{eoverlap}) and the exponential mass hierarchy appears for the charged {leptons}.
	%%%%%%%%%%%%%%%%%%%%%%%%%%%%%%%%%%%%%%%%
	\begin{align}
	m_{11}^{(e)}\ll m_{22}^{(e)}\ll m_{33}^{(e)}.
	\end{align}
	%%%%%%%%%%%%%%%%%%%%%%%%%%%%%%%%%%%%%%%%
On the other hand, this hierarchical structure does not appear in the {neutrinos}. In this model, we can obtain tiny neutrino masses of the order $m_{\nu}\sim{\cal O}(0.1)$eV with putting the value of  the bulk masses as $M_{L}L$, $M_{{\cal N}}L$= ${\cal O}(100)$. In this situation, the immoderate localizations of the {neutrino} zero modes make the effect of the y-dependence of the gauge-singlet VEV $\langle\Phi(y)\rangle$ {weak. Eventually} the non-hierarchical mass structure appears {in} the neutrinos.
	%%%%%%%%%%%%%%%%%%%%%%%%%%%%%%%%%%%%%%%%
	\begin{align}
	m_{11}^{(\nu)}\lesssim m_{22}^{(\nu)}\lesssim m_{33}^{(\nu)}.
	\end{align}
	%%%%%%%%%%%%%%%%%%%%%%%%%%%%%%%%%%%%%%%%

\subsection{Flavor Mixing}
\label{flavormixing}

\begin{figure}[h]
		\centering
		\includegraphics[width=16cm,clip]{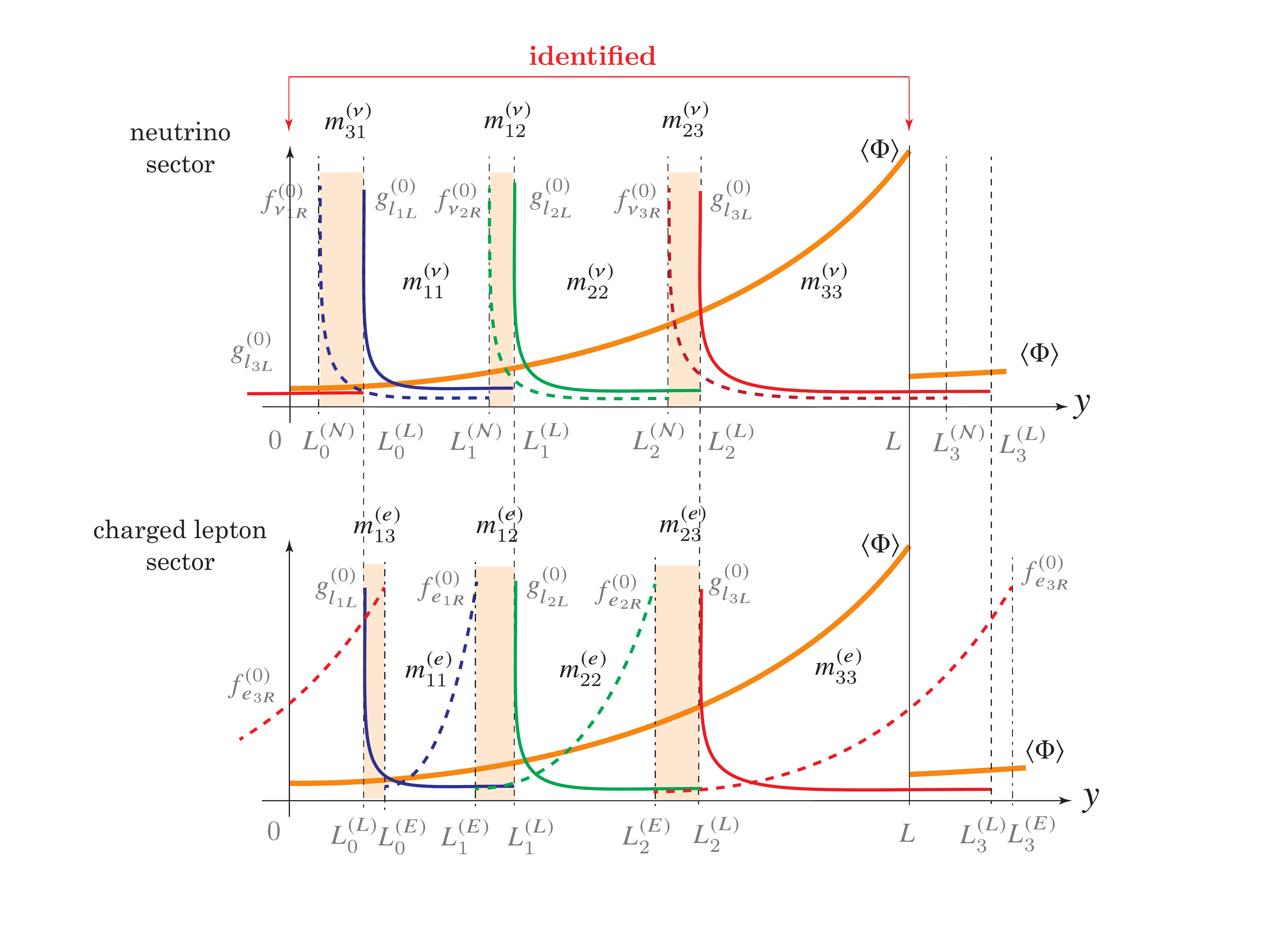}
		\caption{A schematic figure of the correspondence between the components of the mass matrices and the overlap integrals. The orange colored regions indicate the off-diagonal components of the mass matrices.  $m_{ij}^{(\nu)}$ corresponds to the overlap of $g_{l_{iL}}^{(0)}$, $f_{\nu_{jR}}^{(0)}$  and $m_{ij}^{(e)}$ corresponds to {that of} $g_{l_{iL}}^{(0)}$ and $f_{e_{jR}}^{(0)}$.}
		\label{figure-mixing}       % Give a unique label
		\end{figure}

The source of the flavor mixing in this model is the off-diagonal components of the mass matrices~(\ref{nuoverlap})--(\ref{eoverlap}). Since we {introduce} one extra-dimension, the way of the {overlaps} is restricted and we cannot fill up all components of the mass matrices. We found that only six components of the mass matrices are filled up. Under the following suitable choice of the configuration,
	%%%%%%%%%%%%%%%%%%%%%%%%%%%%%%%%%%%%%%%%
	\begin{align}
	&0<L_{0}^{({\cal N})}<L_{0}^{(L)}<L_{1}^{({\cal N})}<L_{1}^{(L)}<L_{2}^{({\cal N})}<L_{2}^{(L)}<L_{3}^{(N)}<L_{3}^{(L)}, \label{nuconfiguration}\\
	&0<L_{0}^{(L)}<L_{0}^{(E)}<L_{1}^{(E)}<L_{1}^{(L)}<L_{2}^{(E)}<L_{2}^{(L)}<L_{3}^{(L)}<L_{3}^{(E)},\label{econfiguration}
	\end{align}
	%%%%%%%%%%%%%%%%%%%%%%%%%%%%%%%%%%%%%%%
the form of the {mass} matrices become {like} the following.
		%%%%%%%%%%%%%%%%%%%%%%%%%%%%%%%%%%%%%%%%
			\begin{align}
			M^{(\nu)}=\left(\begin{array}{ccc}
			m_{11}^{(\nu)}&m_{12}^{(\nu)}&0\\[0.2cm]
			0&m_{22}^{(\nu)}&m_{23}^{(\nu)}\\[0.2cm]
			m_{31}^{(\nu)}&0&m_{33}^{(\nu)}
				\end{array}\right), 
			\hspace{3em}
			M^{(e)}=\left(\begin{array}{ccc}
			m_{11}^{(e)}&m_{12}^{(e)}&m_{13}^{(e)}\\[0.2cm]
			0&m_{22}^{(e)}&m_{23}^{(e)}\\[0.2cm]
			0&0&m_{33}^{(e)}
				\end{array}\right). \label{leptonmassmatrix}
			\end{align}
		%%%%%%%%%%%%%%%%%%%%%%%%%%%%%%%%%%%%%%%%
A schematic figure of the configuration is depicted in Fig.~\ref{figure-mixing}.
		%%%%%%%%%%%%%		figure 	%%%%%%%%%%%%%%%%
		%%%%%%%%%%%%%%%%%%%%%%%%%%%%%%%%%%%%%%%
The above restricted form of the mass matrices is convenient to reproduce the experimental values of the mixing angles and the lepton masses. We found that there is at least one parameter set in this setup, in which we can reproduce the {experimental values with good precision} under the configuration as we will see in the below.

\subsection{CP phase}
\label{CPphase}
Since our model consists of one generation fermion {in 5d}, the Yukawa couplings cannot be a source of the CP phase. Then, the physical CP phase appears from the VEV of the Higgs, which possesses the following y-dependent phase due to the {twisted} BC~(\ref{HBC})~\cite{Fujimoto:2013ki}.
		%%%%%%%%%%%%%%%%%%%%%%%%%%%%%%%%%%%%%%%%
		\begin{align}
		\langle H(y) \rangle &=
							\left( \begin{array}{c}
							0\\
							\langle h(y) \rangle
							\end{array}\right)
							=
							\left( \begin{array}{c}
							0\\
							v/\sqrt{2}
							\end{array}\right)e^{i\frac{\theta}{L}y},\label{Higgs_VEV_form}
		\end{align}
		%%%%%%%%%%%%%%%%%%%%%%%%%%%%%%%%%%%%%%%%
{Through} the overlap integrals~(\ref{nuoverlap})--(\ref{eoverlap}), the VEV of the Higgs {provides a} non-trivial phase to the mass matrix components. This {phase} can be a source of the physical CP phase and {CP violation occurs} in the PMNS matrix. We should emphasize that this Higgs VEV is the only source of the CP phase in our model, which implies that {the magnitude of CP violation in the leptons} can be a prediction after we adjust the CP phase of the quark sector. 

\subsection{Numerical results}
\label{Numerical results}
As a typical example, we choose the parameters as
	%%%%%%%%%%%%%%%%  footnote  %%%%%%%%%%%%%%%%
	\footnote{{The absolute value of the Yukawa couplings $\sqrt{|{\cal Y}^{(\mathcal{N})}|}$, $\sqrt{|{\cal Y}^{(E)}|}$, which are dimension $-1$, are $\sqrt{|{\cal Y}^{(\mathcal{N})}|}=0.00568 L$ and $\sqrt{|{\cal Y}^{(E)}|}=0.0568 L$. Obviously, there is no sizable hierarchical structure for the Yukawa couplings.}}
	 %%%%%%%%%%%%%%%%%%%%%%%%%%%%%%%%%%%%%%%%
 	%%%%%%%%%%%%%%%%%%%%%%%%%%%%%%%%%%%%%%%%
	\begin{align}
	&\tilde{L}_{0}^{(L)}=0.2565,\hspace{2.5em}\tilde{L}_{1}^{(L)}=0.5776, \hspace{2em}\tilde{L}_{2}^{(L)}=0.9432,\nonumber\\
	&\tilde{L}_{0}^{({\cal N})}=0.08240,\hspace{1.6em} \tilde{L}_{1}^{({\cal N})}=0.3909, \hspace{1.8em}\tilde{L}_{2}^{({\cal N})}=0.7317,\nonumber\\
	&\tilde{L}_{0}^{(E)}=0.277, \hspace{2.9em} \tilde{L}_{1}^{(E)}=0.49, \hspace{3em}\tilde{L}_{2}^{(E)}=0.79,\nonumber\\[0.18cm]
	&\tilde{M}_{L}=-136.9,\hspace{1.9em} \tilde{M}_{{\cal N}}=112.1,  \hspace{1.9em}\tilde{M}_{E}=-2.00,\nonumber\\[0.2cm]
	&\tilde{M}_{\Phi}=8.67,\hspace{2.2em}  \tilde{\lambda}_{\Phi}={0.001}, \hspace{2.2em}\frac{1}{\tilde{L}_{+}}=-6.07,  \hspace{1em}\frac{1}{\tilde{L}_{-}}=8.69,\hspace{2.8em} \theta=3, \nonumber\\[0.2cm]
	&{\tilde{{\cal Y}}^{(\mathcal{N})}}=-0.0000309-9.15\times10^{-6} \, i,\hspace{2em} {\tilde{{\cal Y}}^{(E)}}=-0.00309 - 0.000915 \,i
	\end{align}
	%%%%%%%%%%%%%%%%%%%%%%%%%%%%%%%%%%%%%%%%
where the variables with \, $\tilde{}\ $  are dimensionless parameters, which {are scaled} by using the circumference $L$ of the extra dimension.
By calculating the mass eigenvalues, the PMNS matrix and the Jarlskog parameter {in the leptonic sector} $J_{{\rm lepton}}$ through the overlap integrals (\ref{nuoverlap})--(\ref{eoverlap}), the following values {are} obtained.
 	%%%%%%%%%%%%%%%%%%%%%%%%%%%%%%%%%%%%%%%%
	\begin{align}
	&{m_{\nu_1}}=0.0092 \ {\rm eV},\hspace{3.8em} {m_{\nu_2}}=0.013\  {\rm eV}, \hspace{3.4em} {m_{\nu_3}}={0.018}\ {\rm eV},\nonumber\\
	&m_{{\rm electron}}={0.519}\ {\rm MeV},\hspace{1.8em} m_{{\rm muon}}={106}\ {\rm MeV}, \hspace{2.4em}m_{{\rm tau}}={1.778}
	\ {\rm GeV},\nonumber\\[0.2cm]
	&\sin^2\theta_{12}={0.333}, \hspace{2em}\sin^2\theta_{23}={0.435}, \hspace{2em}\sin^2\theta_{13}={0.0239},\nonumber\\
	&J_{{\rm lepton}}={0.0214}\ \ (\sin\delta ={0.607}).
	\end{align}
	%%%%%%%%%%%%%%%%%%%%%%%%%%%%%%%%%%%%%%%%
The ratio {between} the above results and the experimental results are shown in the follows:
 	%%%%%%%%%%%%%%%%%%%%%%%%%%%%%%%%%%%%%%%%
	\begin{align}
	&\sqrt{\frac{\delta m^2}{\delta m^2{}^{({\rm exp.)}}}}={1.03} ,\hspace{2.8em}\sqrt{\frac{\Delta m^2}{\Delta m^2{}^{({\rm exp.)}}}}={0.285},\nonumber\\[0.3cm]
	&\frac{m_{{\rm electron}}}{m_{{\rm electron}}^{({\rm exp.})}}=1.02 ,\hspace{2.8em}\frac{m_{{\rm muon}}}{m_{{\rm muon}}^{({\rm exp.})}}=0.995, \hspace{2.6em}\frac{m_{{\rm tau}}}{m_{{\rm tau}}^{({\rm exp.})}}=1.00,\nonumber\\[0.3cm]
	&\frac{\sin^2\theta_{{\rm 12}}}{\sin^2\theta_{{\rm 12}{}^{({\rm exp.})}}}={1.08} ,\hspace{2.8em}\frac{\sin^2\theta_{{\rm 23}}}{\sin^2\theta_{{\rm 23}{}^{({\rm exp.})}}}={1.02}, \hspace{2.6em}\frac{\sin^2\theta_{{\rm 13}}}{\sin^2\theta_{{\rm 13}{}^{({\rm exp.})}}}={1.02},\label{leptonratio}
	\end{align}
	%%%%%%%%%%%%%%%%%%%%%%%%%%%%%%%%%%%%%%%%
where we defined $\delta m^2$ and $\Delta m^2$ as
 	%%%%%%%%%%%%%%%%%%%%%%%%%%%%%%%%%%%%%%%%
	\begin{align}
	&\delta m^2\equiv m_{\nu_{2}}^2-m_{\nu_{1}}^2,\\[0.15cm]
	&\Delta m^2\equiv m_{\nu_{3}}^2 -\left(\frac{m_{\nu_{1}}^2{+}m_{\nu_{2}}^2}{2}\right),
	\end{align}
	%%%%%%%%%%%%%%%%%%%%%%%%%%%%%%%%%%%%%%%%
according to Ref.~\cite{Capozzi:2013csa}. The mixing angles of the PMNS matrix are within the 3$\sigma$ range \cite{Capozzi:2013csa}. {The value of $\sqrt{\frac{\Delta m^2}{\Delta m^{2({\rm exp.})}}}$ in our example is slightly smaller than experimental range, i.e. about factor-four. An exhaustive parameter scanning would help us to find a rather reasonable point.}

\section{Summary and Discussions}
 We investigated the 5d gauge theory on $S^1$ with point interactions. The {BC's} and the suitable choice of the parameters provide three generations, the charged lepton mass hierarchy, the lepton flavor mixing and tiny degenerated neutrino masses. {The magnitude of the CP violation in the lepton sector} can be a prediction and other physical quantities are reproduced {with good precision} in our model. 
\bibliography{proceedings.bib}

\begin{thebibliography}{9}

\bibitem{Kobayashi:1973fv}
M.~Kobayashi, T.~Maskawa, Prog.Theor.Phys. \textbf{49}, 652 (1973)

\bibitem{Fujimoto:2012wv}
Y.~Fujimoto, T.~Nagasawa, K.~Nishiwaki, M.~Sakamoto, PTEP \textbf{2013}, 023B07
  (2013), \texttt{1209.5150}

\bibitem{Fujimoto:2014fka}
Y.~Fujimoto, K.~Nishiwaki, M.~Sakamoto, R.~Takahashi, JHEP \textbf{1410}, 191
  (2014), \texttt{1405.5872}

\bibitem{Reed:1975uy}
M.~Reed, B.~Simon, Methods of Modern Mathematical Physics. 2. Fourier Analysis,
  Selfadjointness  (1975)

\bibitem{Seva1986}
P.~Seva, J. of Phys \textbf{36}, 667 (1986)

\bibitem{Cheon:2000tq}
T.~Cheon, T.~Fulop, I.~Tsutsui, Annals Phys. \textbf{294}, 1 (2001),
  \texttt{quant-ph/0008123}

\bibitem{Fujimoto:2013ki}
Y.~Fujimoto, K.~Nishiwaki, M.~Sakamoto, Phys.Rev. \textbf{D88}, 115007 (2013),
  \texttt{1301.7253}

\bibitem{Fujimoto:2011kf}
Y.~Fujimoto, T.~Nagasawa, S.~Ohya, M.~Sakamoto, Prog.Theor.Phys. \textbf{126},
  841 (2011), \texttt{1108.1976}

\bibitem{Capozzi:2013csa}
F.~Capozzi, G.~Fogli, E.~Lisi, A.~Marrone, D.~Montanino et~al., Phys.Rev.
  \textbf{D89}, 093018 (2014), \texttt{1312.2878}

\end{thebibliography}

\end{document}